\documentclass[11pt]{article} 

\usepackage{helvet}

\usepackage{graphicx} 
\usepackage{mathtools}
\usepackage{amsmath}
\usepackage{amsfonts}
\usepackage{multirow}
\usepackage[export]{adjustbox}
\usepackage[left=1in, right=1in, top=1in, bottom=1in]{geometry} 
\usepackage{booktabs}
\usepackage{rotating}
\usepackage{rotfloat}
\usepackage{float}
\usepackage{subfig}
\usepackage{url}
\usepackage{pdfpages}
\usepackage{rotating}
\usepackage[doublespacing]{setspace} 
\usepackage{csvsimple}
\usepackage{array}
\usepackage{wrapfig}
\usepackage[utf8]{inputenc}

\usepackage{cite}
\usepackage{tabularx}
\usepackage{float}

\usepackage{amsthm}                
\usepackage{amssymb}

\usepackage{multirow}

\usepackage{caption}

\usepackage{amsfonts}
\usepackage{amsmath} 
\usepackage{tikz}
\usetikzlibrary{positioning,shapes.geometric}

\usepackage{xcolor}

\usepackage{setspace}

\usepackage{mathrsfs}

\makeatletter
\newcommand*{\indep}{%
  \mathbin{%
    \mathpalette{\@indep}{}%
  }%
}
\newcommand*{\nindep}{%
  \mathbin{
    \mathpalette{\@indep}{\not}
  }%
}
\newcommand*{\@indep}[2]{%
  \sbox0{$#1\perp\m@th$}
  \sbox2{$#1=$}
  \sbox4{$#1\vcenter{}$}
  \rlap{\copy0}
  \dimen@=\dimexpr\ht2-\ht4-.2pt\relax
  \kern\dimen@
  {#2}%
  \kern\dimen@
  \copy0 
} 
\makeatother

\usepackage{xr-hyper}

\definecolor{forestgreen}{RGB}{34,139,34}
\usepackage{hyperref}
\hypersetup{
    colorlinks=true,
    linkcolor=magenta,
    filecolor=magenta, 
    citecolor=forestgreen,      
    urlcolor=blue
}

\DeclareFontFamily{U}{mathx}{\hyphenchar\font45}
\DeclareFontShape{U}{mathx}{m}{n}{
      <5> <6> <7> <8> <9> <10>
      <10.95> <12> <14.4> <17.28> <20.74> <24.88>
      mathx10
      }{}
\DeclareSymbolFont{mathx}{U}{mathx}{m}{n}
\DeclareFontSubstitution{U}{mathx}{m}{n}
\DeclareMathAccent{\widecheck}{0}{mathx}{"71}
\DeclareMathAccent{\wideparen}{0}{mathx}{"75}

\usepackage{sectsty}
\sectionfont{\large}

\usepackage{layout}

\usepackage{geometry}

\newcolumntype{C}[1]{>{\centering\arraybackslash}p{#1}}

\usepackage{authblk}

\usepackage{afterpage}
\usepackage{lscape}
\usepackage{graphicx}
\usepackage{amsmath}
\usepackage{relsize} 

\usepackage{datetime}
\def\paperversionmajor{5}
\def\paperversionminor{0}

\makeatletter
\newcommand*{\addFileDependency}[1]{
  \typeout{(#1)}
  \@addtofilelist{#1}
  \IfFileExists{#1}{}{\typeout{No file #1.}}
}
\makeatother

\makeatletter
\newcommand{\vast}{\bBigg@{3}}
\newcommand{\Vast}{\bBigg@{4}}
\makeatother

\begin{document}

\title{Transportability methods for failure-time outcomes under assumptions for relative effect measures}

\author[1]{Sarah E. Robertson}
\author[2]{Lan Wen}
\author[3,4]{Lawson Ung}
\author[5]{Jon A. Steingrimsson}
\author[3,4,6]{Issa J. Dahabreh}

\affil[1]{Department of Epidemiology, Dartmouth College, Hanover, NH}
\affil[2]{Department of Statistics and Actuarial Science, University of Waterloo, ON, Canada}
\affil[3]{CAUSALab, Harvard T.H. Chan School of Public Health, Boston, MA}
\affil[4]{Department of Epidemiology, Harvard T.H. Chan School of Public Health, Boston, MA}
\affil[5]{Department of Biostatistics, Brown University School of Public Health, Providence, RI }
\affil[6]{Department of Biostatistics, Harvard T.H. Chan School of Public Health, Boston, MA}

\maketitle{}
\thispagestyle{empty} 

\thispagestyle{empty}

\clearpage
\thispagestyle{empty} 
\vspace*{0.75in}
\begin{abstract}
\noindent
\linespread{1.3}\selectfont 

 In certain clinical areas, relative effect measures are believed to remain more constant across populations than absolute measures. This suggests that exchangeability assumptions on the relative scale may be more plausible than those on the absolute scale. Despite this, most methods for extending trial results to a target population rely on strong and often implausible distributional exchangeability assumptions between the trial and target population (distributional transportability). 
 
 We propose identification results for failure-time outcomes under a more plausible assumption of exchangeability on the relative scale, focusing primarily on the risk ratio (risk ratio transportability).
We applied these methods to estimate the effect of screening strategies on all-cause mortality, using data from the National Lung Screening Trial (NLST) to a nationally representative target population from the National Health Interview Survey (NHIS). All-cause mortality data was available from NHIS, along with no confounding of screening, allowing a falsification assessment of the distributional transportability analysis against the observed baseline risk. 

We found that this approach substantially underestimated mortality risk. In comparison, the risk ratio transportability approach produced reasonable estimates that matched the baseline risk. 
 Thus, assumptions on relative scales may be more plausible in certain settings where distributional transportability fails.

\end{abstract}

\clearpage
 \pagenumbering{arabic}
\setcounter{page}{1}
\section*{INTRODUCTION}

Transportability and generalizability analyses aim to extend results from a trial to a target population \cite{cole2010, westreich2017, rudolph2017, dahabreh2018generalizing, dahabreh2020transportingStatMed}. These analyses depend on exchangeability conditions that require adjustment for baseline covariates to ensure that the trial and target population are conditionally exchangeable. The most commonly invoked assumptions require
independence between either the entire distribution of potential outcome and participation, given covariates (\textit{distributional transportability}) or in expectation 
(\textit{mean transportability}). In many settings, these assumptions may be implausible. 

Instead, recent work has proposed working under weaker, scale-dependent assumptions, such as exchangeability on the relative scale (e.g., risk ratio transportability) \cite{schwartz2006ratio, kent2016risk, huitfeldt2018choice, harrell2020}. These assumptions have been developed for cross-sectional outcomes \cite{dahabreh2022learning}. This paper proposes scale-dependent assumptions for failure-time outcomes, which generally are limited to invoking only distributional transportability assumptions \cite{cole2010, lu2019generalizing, zuo2022transportability,lee2022transporting}. 
 
 Here, we derive identification results for analyses under risk ratio transportability to learn about causal effects for failure-time outcomes in the target population \cite{dahabreh2022learning}. We applied these methods to estimate the effect of screening strategies on all-cause mortality, extending findings from the National Lung Screening Trial (NLST) \cite{national2011reduced} to a nationally representative target population from the National Health Interview Survey (NHIS). All-cause mortality data was available from NHIS, along with no confounding of screening, allowing a falsification assessment of the distributional transportability analysis against the observed baseline risk. 

\section*{STUDY DESIGN, INFORMATION SOURCES, AND CAUSAL EFFECTS IN FAILURE-TIME TRANSPORTABILITY ANALYSES}

\paragraph{Study design:} Transportability analyses often use a non-nested trial design, where trial data is combined with a separately obtained sample from a target population. Under this design, the trial and target population samples are obtained independently from their underlying population of screening-eligible individuals. For example, the NLST combined with the NHIS has a non-nested design, where the NHIS is drawn under a complex survey design that we account for using survey weights \cite{dahabreh2021studydesigns}. 

\paragraph{Trial (NLST):} The NLST \cite{national2011reduced, national2013results} enrolled participants at 33 centers in the United States from August 2002 through April 2004. Eligible individuals were 55 through 74 years-old, current or former smokers with at least 30 pack-years of smoking. Former smokers were required to have quit within 15 years of enrollment. Individuals were excluded if they had received a low-dose CT chest scan in the past 18 months, had a history of lung cancer or history of non-lung cancer within the past 5 years, had recent medical problems, participated in other studies of cancer screening, had any portion of their lungs removed, were unable to lie flat with arms raised overhead, required home oxygen supplementation, or had metallic implants in the chest or back. Trial participants were randomized to low-dose CT or chest radiography screening, with screening assessments performed at baseline, then yearly for two more years. Participants were followed for events yearly through questionnaires or National Death Index searches. We obtained data for the NLST from the National Cancer Institute (NCI) \cite{nciwebsite} and used the follow-up time data as recommended in the NCI's NLST User Guide \cite{nciwebsite}. As in the original NLST analyses, we used an event cutoff date of December 31, 2009 for all-cause mortality \cite{national2011reduced}. 

\paragraph{Target population (NHIS):} NHIS uses a multistage cluster sample design to select households representative of the civilian non-institutionalized United States population \cite{parsons2014design, nhisCDCweb}. NHIS is run every year, but only some years include the Cancer Control Supplement, which has detailed information on smoking history for former smokers \cite{nhisCCSweb}. We used 2005 NHIS data because that year included the Cancer Control Supplement and was closest to the start date of the NLST. We also selected this year because during the 6-year follow-up time of interest (2005 through 2011), low-dose CT screening for lung cancer was not yet readily available, predating the 2013 USPSTF recommendation \cite{doria2012use, jemal2017lung}. A later NHIS dataset would have follow-up overlapping with the introduction of CT screening (e.g., 2010 through 2016). We accessed NHIS data through the IPUMS USA database \cite{NHISdata}. NHIS data are linked to the National Death Index for all cause death with follow-up through December 31, 2019. For our analysis, follow-up time was calculated from the 2005 interview date and administratively censored at six years to be consistent with the primary NLST endpoint. We had access to the public-use mortality files, which provided the quarter of death and the year. We measured survival time from the NHIS interview to death  or administrative censoring at six years.

We restricted the NHIS sample to NLST-eligible individuals. More specifically, we restricted the NHIS sample to ever-smokers (having smoked at least 100 cigarettes in their lifetime) that completed the supplemental survey questions regarding smoking history. Next, we applied the NLST inclusion criteria of age and smoking history to the NHIS sample. We used the same definition of pack-years as the NLST to calculate the pack-year history in NHIS \cite{NHISdatadicctionary}. We also excluded individuals who had received cancer treatment in the past year, had any non-melanoma cancer diagnosis in the past five years, were unable to reach overhead or had limited mobility, or were unable to consent to trial participation. Full details of the implementation is available elsewhere \cite{robertson2024comparing}. Web Appendix 1 has a detailed flow chart of the exclusion criteria. 

\paragraph{Composite dataset:} To form the composite dataset, we appended the individual patient data from the NLST with the NHIS. When forming the composite dataset, we selected baseline covariates that were measured similarly in both datasets, such as demographic information (age, gender, body mass index, race, marital status, and education), smoking history (smoking status and pack years), and comorbidities (diabetes, emphysema, heart disease, hypertension, stroke) used in prior work \cite{robertson2024comparing}. 

\paragraph{Causal effects in failure-time transportability analyses:} We adopted a discrete-time framework for failure-time outcomes \cite{hernan2000marginal, tutz2016modeling}, discretizing the follow-up time into non-overlapping quarterly intervals up to six years. When examining failure-time outcomes, the potential (counterfactual) outcome \cite{Rubin1974, robins2000d} of interest is the outcome at a follow-up interval of interest under a joint intervention that assigns a time-fixed screening strategy and eliminates right-censoring. Specifically, our estimand is the effect of assignment to these screening strategies, allowing for non-adherence (i.e., the intention-to-treat effect) \cite{dahabreh2019identification}. 
This analysis relies on the assumption that there are no unmeasured common causes of adherence and participation in the NLST, and that participation does not affect adherence directly \cite{dahabreh2022generalizing}. In Web Appendix 2 we provide a more formal description of the setup, including notation.

The specific strategies compared are assignment to three annual low-dose CT screening rounds versus assignment to a control strategy where CT screening is absent. In our analysis, the control strategy is defined by three annual chest radiography screening rounds in the NLST and by no screening for lung cancer in the NHIS population. We group these control strategies together under the assumption that both have similar effects on all-cause mortality. We check this assumption in a falsification assessment and it is supported by the  Prostate, Lung, Colorectal, and Ovarian Cancer Screening Trial (PLCO) trial, which found no benefit on mortality when comparing these two strategies. In Web Appendix 13, we relax this assumption by using PLCO data and find that our results are similar to the main analysis, supporting the assumption. We are primarily interested in comparing the risk (cumulative incidence) between the screening strategies, at different time-points up to six years in the target NHIS population. We are also interested in comparing the risks, risk difference, and risk ratio at six years.

\section*{STATISTICAL ANALYSIS} 

\paragraph{Comparing different transportability assumptions:} To learn about causal effects in the target population, we assumed exchangeability, conditional on baseline covariates, between the NLST participants and NHIS target population. More specifically, we assumed conditional exchangeability on the risk ratio scale at each follow-up interval (``risk ratio transportability'') between the trial and target population. Risk ratio transportability is typically more plausible than the commonly invoked distributional transportability assumption because effect heterogeneity is scale-dependent. Background knowledge suggests that relative effects remain stable across populations, even when baseline risks vary. For distributional transportability to hold, we must adjust for all covariates that affect all-cause mortality and have a different distribution among NLST and NHIS participants. While for risk ratio transportability to hold, we must adjust for covariates that modify the effect on the relative scale, which is a much smaller subset.

In addition to the main transportability assumptions, identifying causal effects in the target populations requires additional assumptions. We rely on standard assumptions underlying trial-only analyses (e.g., no unobserved confounding due to randomization; \cite{hernancausal}). We also require positivity over trial participation status; that is, we require every pattern of covariates relevant for transportability in the target population to be observed with nonzero probability in the trial.

Identifying causal effects under risk ratio transportability requires applying the relative effect measure from the trial to the baseline risk of the target population. In contrast, identifying causal effects under distributional transportability only uses trial data and does not require information from the target population. As a result, only when working under risk ratio transportability do we require an additional assumption of no unmeasured confounding for the control screening strategy in the target population (a partial exchangeability assumption). 
 In NHIS, because individuals could not receive low-dose CT or chest radiography  screening, and we can assume there is uniform use of a control screening strategy (i.e., no screening), we expect partial exchangeability to hold (see Wen and Sarvet \cite{wen2025estimating} for further discussion when not all treatments are available in the target population). In Web Appendices 3 and 4 we more formally list these conditions and provide the proof for identifying the counterfactual survival under the screening strategies in the target population.

Given our focus on screening in our application, we used the risk ratio scale throughout for the exchangeability assumption, since it is suggested to use this scale for strategies that reduce incidence (are protective) \cite{huitfeldt2019effect, sheps1958shall}. It is also possible to use a transportability assumption for other relative effect measures besides the risk ratio, such as the survival ratio or hazard ratio. We provide details for working on the survival ratio and hazard ratio scale in Web Appendices 6 and 7. In Web Appendix 8 we provide a more formal comparison of these transportability assumptions to show that it is unlikely that all hold at the same time. 

\paragraph{Transportability analyses:} As our primary analysis, we estimated the risk curves for the screening strategies in the target population under risk ratio transportability. As a falsification assessment, we compared the risk curves for the control screening strategy estimated under distributional transportability against the curves estimated from the NHIS-only and NLST-only data. We also compared results at six years under both risk ratio transportability and distributional transportability with the results from the NLST-only analysis. 

We describe the approaches we used below for risk ratio transportability and distributional transportability. Note that for the estimator under risk ratio transportability that the risk under low-dose CT screening is transported, while the risk under the control group is fit using the NHIS-only data directly (without a transportability formula). In contrast, when using the estimator under distributional transportability, both screening strategies must be transported, because this estimator does not use screening or outcome data from NHIS.
Throughout, we used a bootstrap procedure recommended and appropriate for survey data, to obtain 95\% confidence intervals \cite{li2023estimating, rao1988resampling,shao2003impact}. In this application, we obtained results using 1000 bootstrap runs that account for the survey design of the NHIS. Full details for estimation and inference using these approaches are described in Web Appendix 5.
 
\paragraph{Estimating the risk curve for the low-dose CT screening strategy in the target population under risk ratio transportability:} 
The risk ratio transportability approach transports inferences from the NLST to NHIS by multiplying estimates of the conditional baseline risk fit in NHIS with the conditional risk ratio fit in the NLST, then marginalizing these estimates over only the NHIS individuals to obtain the risk for the low-dose CT screening strategy in the NHIS target population. To account for the survey design of NHIS, we used
survey weights to calculate the weighted mean of the conditional predicted risks when marginalizing over the covariate distribution of the target population.

We estimated the conditional baseline risk in NHIS by fitting a pooled logistic regression model to estimate the discrete-time conditional hazards. We then used the product limit estimator to obtain the conditional survival probabilities, with the risk as their complement. 

Similarly, to estimate the conditional risk ratio in the NLST, we estimated the discrete-time conditional hazards using pooled logistic regression models fit separately in the low-dose CT and control screening arms. We used the product limit estimator to obtain the conditional survival probabilities in each arm (with the risk calculated as the complement), and calculated the ratio of these risks.

In both NHIS and NLST, the pooled logistic regression models were conditional on baseline covariates, time, and remaining in the risk set of the outcome by that interval \cite{hernancausal}. We included the main effects of all the selected baseline covariates (demographic, smoking history, and comorbidities) and used restricted cubic splines with nine knots for time. We placed the knots to allow flexibility early on (every quarter up to 3) and every third quarter up to 21 quarters. We used survey weights when fitting the pooled logistic regression model in NHIS. 

Furthermore, while the risk ratio transportability approach is not guaranteed to produce monotonically decreasing survival curves (or equivalently, monotonically increasing risk curves) this did not occur in our application, but we describe post-hoc corrections \cite{denz2023comparison} (e.g., truncation or isotonic regression) in the Appendix.
Last, in Web Appendix 10, we present results for estimators that are valid under the alternative assumptions of survival ratio transportability and hazard ratio transportability.

\paragraph{Estimating the risk curve for the control screening strategy in the target population under risk ratio transportability:} 
We have failure-time outcome information in NHIS. To meet the partial exchangeability assumption, we assume there is no unmeasured confounding of the control screening strategy in NHIS. This is supported because low-dose CT screening for lung cancer was not readily available during the 6-year follow-up period (from 2005-2011). Therefore, we do not need to use a transportability formula to estimate the effect under the control screening strategy in NHIS. We used a g-formula (outcome-modeling) estimator fit in the NHIS data \cite{dahabreh2020transportingStatMed}. We used the baseline risk model fit in NHIS as described above, and marginalized the conditional risk estimates  over the covariate distribution of the NHIS participants. 

\paragraph{Estimating the risk curves for both screening strategies in the target population under distributional transportability:}
Under the assumption of distributional transportability we transport inferences from the NLST to NHIS by marginalizing estimates of the conditional risk fit in the NLST, over the NHIS participants (using a weighted mean to take the survey design into account). We used the same estimates of the risk fit in the NLST as the estimator under risk ratio transportability (using the pooled logistic regression models fit separately in each screening strategy for the NLST participants). 

\paragraph{Falsification assessment for the control screening strategy:} To  evaluate the validity of the distributional transportability analysis, which does not rely on outcome data from the target population, we used a falsification assessment. We compared the predicted risk curve for the control screening strategy, derived under the distributional transportability assumption, against the observed risk curve in NHIS. Similarity between these curves supports that the assumption holds.
For the validation assessment to be valid, we assume consistency of no screening and radiography screening jointly as control screening strategies (i.e., no treatment variation). If the assumption of distributional transportability is supported for the control strategy, we may expect it to hold for the low-dose CT screening strategy as well. 


\section*{RESULTS} 
We restricted our analysis to NLST participants with complete information, conditional on the selected baseline covariates that could be measured similarly in both datasets, leaving us with a sample size of 51,274 NLST participants. Missingness was limited across these covariates in the NLST (4.1\% of NLST observations). In our target population sample for NHIS, after restricting to the covariate eligibility criteria, 1018 individuals remained; when restricting to individuals with complete-cases on the selected covariates of interest, we were left with a sample size of 992. In total, 41 individuals were ineligible to be linked to the National Death Index, leaving us with a sample size of 951 NHIS participants. Web Figure 1 shows a flowchart of the NHIS data and Web Table 1 summarizes the distribution of baseline covariates for both the NLST and NHIS samples. Censoring was limited in both the NLST and NHIS. Over six years of follow-up time, in the NLST, 3265 participants died and 7836 were censored (out of a sample size of 51,274). In NHIS, 153 participants died and 0 were censored (out of a sample size of 951).

\begin{figure}[!ht]
    \centering
    \caption{Risk curves in the target population (NHIS) under risk ratio transportability.}
    \label{fig:1_new_method}
    \includegraphics[width=10cm]{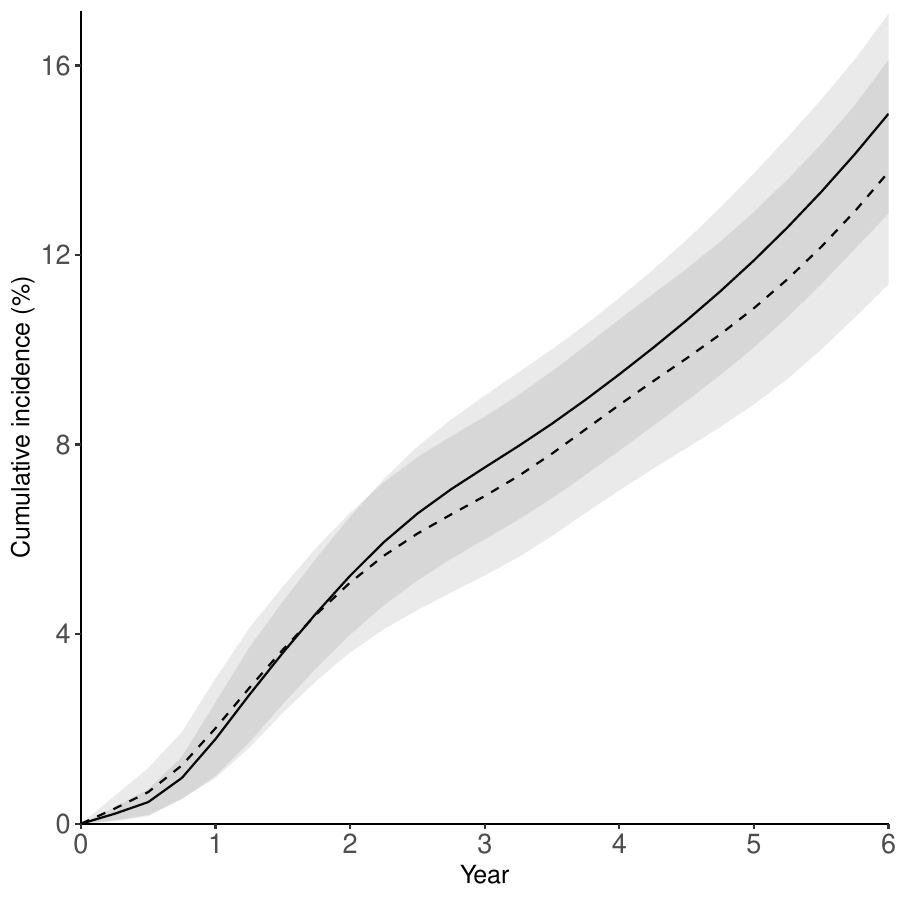}
    \caption*{Black lines indicate point estimates and grey bands indicate 95\% confidence intervals (CI). \\\
    Short dashed line = low-dose CT (with CI indicated by the dark grey band);
    Solid line = control (with CI indicated by the light grey band). }
\end{figure}

\begin{figure}[!ht]
    \centering
    \caption{Risk difference in the target population (NHIS) under risk ratio transportability.}
    \label{fig:2_new_method}
    \includegraphics[width=10cm]{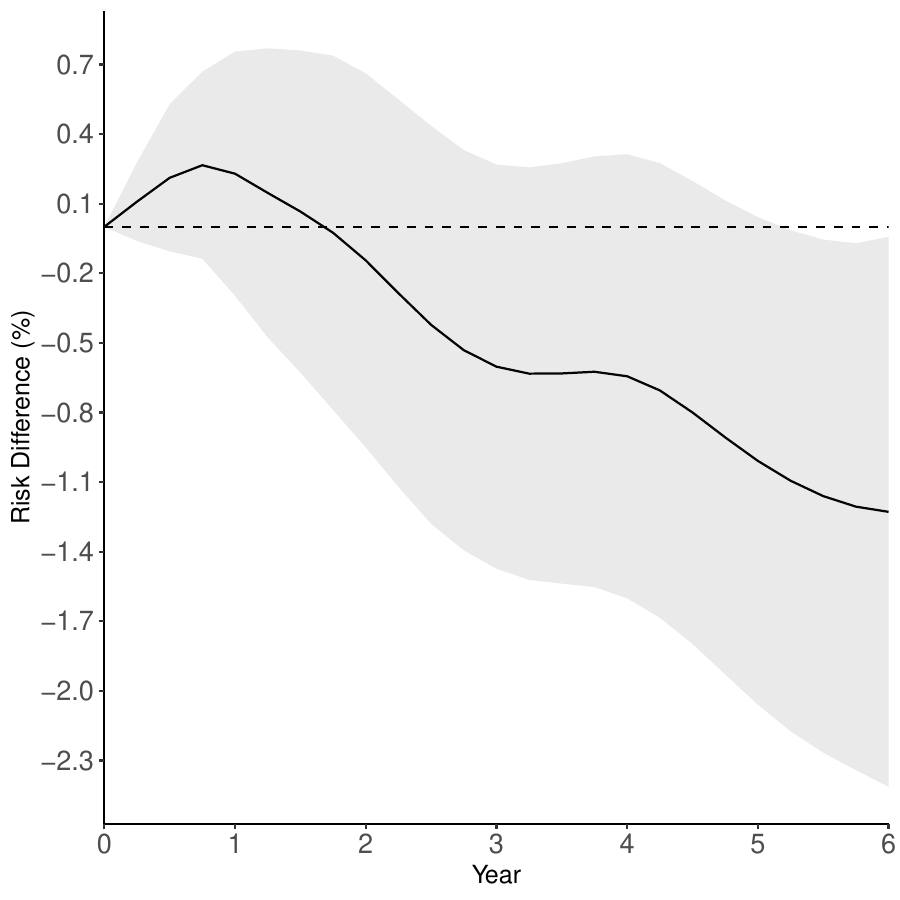}
   \caption*{Black lines indicate point estimates and grey bands indicate 95\% confidence intervals (CI). \\\
    }
\end{figure}

\paragraph{Estimated risk curves under risk ratio transportability:}
Figure \ref{fig:1_new_method} shows the estimated risk curves for the target population under risk ratio transportability. After the first two years, the risk associated with low-dose CT screening is lower than that of the control strategy, although the confidence intervals remain wide and overlap. Figure \ref{fig:2_new_method} shows the corresponding risk difference curve for the target population between the screening strategies. There is a benefit of the low-dose CT screening strategy starting at five years, when the 95\% confidence interval no longer crosses zero. At six years, the risk difference for low-dose CT screening versus control screening is -1.23\%. In Web Appendix 9 we explain why the results for the risk difference are more precise than the risk curves suggest, due to correlation between the estimated risk in the low-dose CT screening strategy and the control screening strategy when using the estimator under risk ratio transportability.

\begin{figure}[!ht]
    \centering
    \caption{Risk curves under the control screening strategy.}
    \label{fig:1_new_A0_transport}\includegraphics[width=10cm]{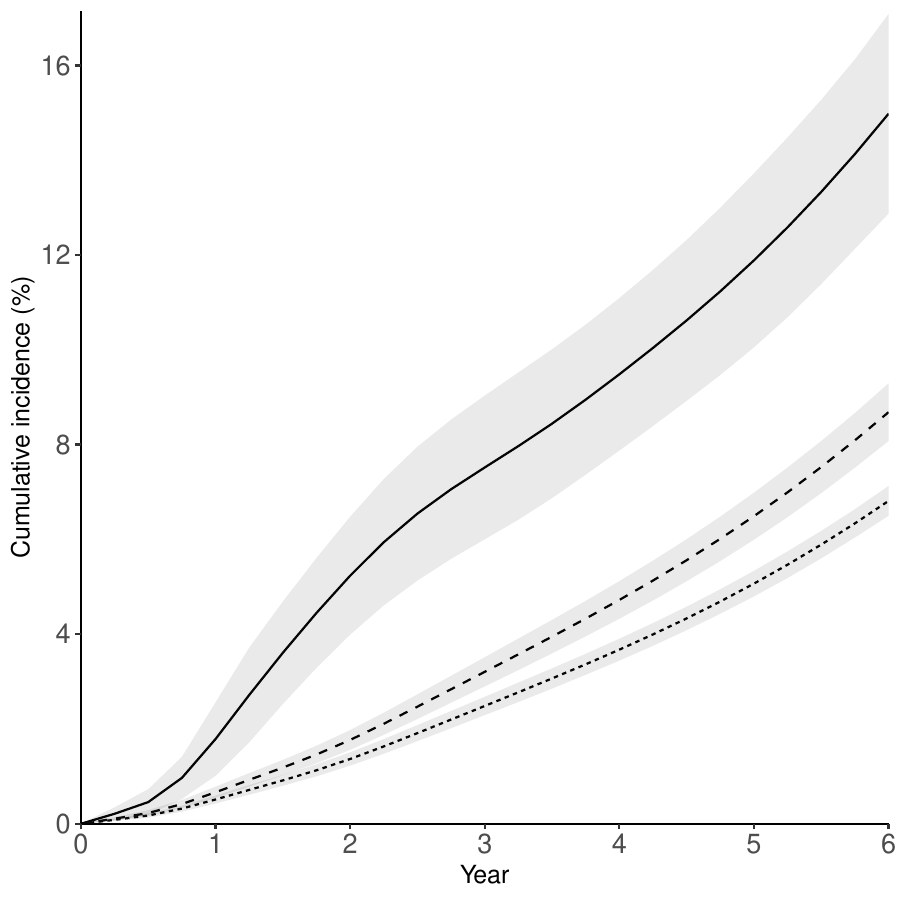}
    \caption*{Black lines indicate point estimates and grey bands indicate 95\% confidence intervals. \\\
    Note that the black solid line is the same as that shown in Figure 1. \\\ 
     Solid line = Target population (NHIS-only); Long dashed line = Distributional transportability; Short dashed line = NLST-only. 
   } 
\end{figure}

\paragraph{Validation assessment of the risk curve for the control screening strategy:}
Figure \ref{fig:1_new_A0_transport} shows
estimated risk curves for the control screening strategy only, plotting an NLST-only curve, NHIS-only curve, and distributional transportability risk curve. The NHIS-only risk curve is higher than the NLST-only risk curve, while the distributional transportability risk curve falls between the two.  Although the distributional transportability risk curve approximates the risk in the NHIS more closely than the NLST-only curve, it still underestimates the risk in the NHIS population - for example, by about 7\% at six years. Additionally, the confidence intervals do not overlap between the NHIS-only and distributional transportability curve beyond the first year, suggesting the estimates are significantly different. The comparison between the NHIS-only curve and distributional transportability curve allows for an informal validation check of the transportability method; since they are different, this suggests that the transportability analysis may not be reasonable due to a violation in assumptions.

This difference is large enough that it cannot be explained by a difference in the control interventions, and instead suggests a violation in assumptions. The NHIS-only population were under no screening strategy (and unlikely to receive other screening strategies due to the time period we selected), while the control screening strategy in the NLST-only population were assigned to radiography. However, we expect radiography screening to have a similar effect on all-cause mortality as no screening, based on evidence from the Prostate, Lung, Colorectal, and Ovarian Cancer Screening Trial (PLCO) trial which found no benefit when comparing these two screening strategies \cite{oken2011screening}. Thus, the estimator under distributional transportability is not reasonable, since it does not approximate the NHIS-only curve.

\paragraph{Comparison of results at six years:}
Table \ref{tab2} compares the risks, risk differences, and risk ratios for the NLST-only and transportability analyses at six years. The risk ratio transportability analysis directly assesses the empirical baseline risk under the control screening strategy in the national population, using the NHIS-only data, and estimates a risk of 15.0\%. In contrast, both the NLST-only and distributional transportability analyses underestimate the risk in the target NHIS population under the control screening strategy, with estimates of 6.3\% and 8.0\%, respectively. These analyses also appear to underestimate the risk for the low-dose CT screening strategy in the target NHIS population, although this cannot be directly evaluated because the target population is unscreened.

Instead, we indirectly examined whether the risk difference between the estimated risk for the low-dose CT screening strategy for each method (NLST-only and distributional transportability), anchored against the 15.0\% estimated risk for the control screening strategy from the NHIS-only data, is plausible. As shown in Table \ref{tab_differences}, the calculated difference in the risk differences for the NLST-only and distributional transportability methods were approximately 8\% and 6\%, respectively- both highly implausible effect sizes for a screening strategy over a relatively short period, given that the NLST found a risk difference of only -0.52\% between the screening strategies. In contrast, the estimator under risk ratio transportability provided a more reasonable risk difference of -1.29\%, which falls within the confidence interval of the NLST-only analysis.


\section*{DISCUSSION}
We relaxed the commonly invoked assumption of distributional transportability for failure-time outcomes by working under the weaker assumption of risk ratio transportability, and proposed estimators for screening effects in the target population. Using the NLST and NHIS data, we found that distributional transportability failed the validation assessment, underestimating the six-year all-cause mortality risk by about 7\%. A similar failure occurred when emulating a colorectal screening trial for all-cause mortality, which found an impossibly significant benefit due to strong unmeasured confounding \cite{garcia2017effectiveness, garcia2019estimating}. In contrast, the risk ratio transportability approach resulted in estimates anchored locally to the estimated baseline risk in NHIS.  

We also found that the risk ratio was in close agreement across the transportability methods, suggesting relative effect measures may be less sensitive to the choice of transportability assumption. Therefore, the risk ratio transportability assumption may be most important to invoke when estimating the risks or absolute effects such as the risk difference, where anchoring on the correct baseline risk is critical.

The decision for which transportability assumption to use, which leads to different methods, should be made a priori to the analysis, based on substantive knowledge. The estimator under distributional transportability only requires baseline covariate data from the target population and does not require assuming the absence of unmeasured confounding for the control strategy. In contrast, the estimator under risk ratio transportability must either have outcome information available or have a reasonable baseline risk model. It also must have no confounding in the target population, which is only guaranteed to hold in settings where there is no treatment variation, as in our application where CT screening was not routinely available. For either method, sensitivity analyses that examine violations of the transportability assumptions should be considered \cite{dahabreh2023sensitivity}.

\clearpage
\section*{Figures}

\clearpage
\section*{Tables}

\begin{table}[!ht]
\caption{Results at six years for the risks, risk difference, and risk ratio.}\label{tab2}
\begin{tabular}{lllll}
\toprule 
Analysis & Low-dose CT  (\%)  & Control (\%)    & Difference (\%)    &  Ratio          \\ \midrule
Trial-only    & 6.3 (6.0, 6.6)  & 6.8 (6.5, 7.1) & -0.53 (-0.94, -0.11)  & 0.92 (0.86, 0.98)  \\ \midrule
 \multirow{2}{*}{\parbox{4.5cm}{Target population:}} &   &   &   & \\
   &  &  &  &  \\ 
 \multirow{2}{*}{\parbox{4.2cm}{Risk ratio \\ transportability}} & 13.7  (11.4, 16.1)    & 15.0 (12.9,17.1)$^*$  & -1.23 (-2.41, -0.04)  & 0.92 (0.84, 1.00)   \\ 
   &  &  &  &  \\ 
      &  &  &  &  \\ 
\multirow{2}{*}{\parbox{4.2cm}
    {Distributional \\ transportability}} &  8.0 (7.4, 8.6) & 8.7 (8.1, 9.3) & -0.62 (-1.23, -0.01)  & 0.92 (0.79, 1.04)   \\ 
   &  &  &  &  \\ 
\bottomrule
\end{tabular}
\caption*{95\% confidence intervals are from 1000 bootstrap runs. \\
$^*$  Note this is estimated directly using the NHIS-only data.} 
\end{table}

\begin{table}[!ht]
\caption{Differences between analyses for the risks and risk difference. Ratios between analyses for the relative risk (the relative risk ratio).}\label{tab_differences}
\begin{tabular}{lllll}
\toprule 
Comparison & Low-dose CT  (\%)  & Control (\%)    & Difference (\%)    &  Ratio          \\ \midrule
\multirow{2}{*}{\parbox{5cm}{Risk ratio transportability vs.:}} &   &   &   & \\
   &  &  &  &  \\ 
 \hspace{0.25cm} Distributional transportability  &  5.70 (3.48, 7.90) & 6.30 (4.21,  8.38) & -0.61 (-1.34, 0.22)  & 0.99 (0.94, 1.03)   \\
 \hspace{0.25cm} Trial-only   & 7.47 (5.16, 9.77)  & 8.17 (6.03,10.31) & -0.70 (-1.62,0.22)  & 0.99 (0.94,1.05)  \\ \bottomrule
\end{tabular}
\caption*{95\% confidence intervals are from 1000 bootstrap runs. 
} 
\end{table}

\clearpage
\bibliographystyle{ieeetr}
\bibliography{bibliography_FT_NLST}


\ddmmyyyydate 
\newtimeformat{24h60m60s}{\twodigit{\THEHOUR}.\twodigit{\THEMINUTE}.32}
\settimeformat{24h60m60s}
\begin{center}
\vspace{\fill}\ \newline
\textcolor{black}{{\tiny $ $generalizability\_relative\_measures, $ $ }
{\tiny $ $Date: \today~~ \currenttime $ $ }
{\tiny $ $Version: \paperversionmajor.\paperversionminor $ $ }}
\end{center}

\end{document}